\documentclass[12pt]{amsart}
\usepackage{geometry}            
\usepackage{graphicx}
\usepackage{amssymb}
\usepackage{mathtools}
\usepackage{epstopdf}
\usepackage{color}
\usepackage{mathpazo}
\fontfamily{Palatino}
\newcommand{\be}{\begin{equation}}
\newcommand{\ee}{\end{equation}}
\newcommand{\ba}{\begin{eqnarray}}
\newcommand{\ea}{\end{eqnarray}}

\begin{document}

\begin{center}
\tiny{Essay written for the Gravity Research Foundation 2021
Awards for Essays on Gravitation, Honorable Mention}

\vspace{1.5cm}
\huge{\bf Through a black hole into a New Universe}
\vspace{1cm}

\large{\bf  Robert Brandenberger$^*$, Lavinia Heisenberg$^\dagger$ and Jakob Robnik$^\dagger$}  \\
\vspace{1cm}
\end{center}
{\tiny{
$^*$Department of Physics, McGill University, Montr\'{e}al, QC, H3A 2T8, Canada.\\
$^\dagger$ Institute for Theoretical Physics, ETH Z\"urich, Wolfgang-Pauli-Strasse 27, 8093,  Z\"urich, Switzerland.\\
\vspace{1cm}
\begin{center}
{\large{\bf Abstract}}
\end{center}
\small{We show that an S-Brane which arises in the inside of the black hole horizon when the Weyl curvature reaches the string scale induces a continuous transition between the inside of the black hole and the beginning of a new universe. This provides a simultaneous resolution of both the black hole and Big Bang singularities. In this context, the black hole information loss problem is also naturally resolved.}
\vspace{1.5cm}
\newline
email for RB: rhb@physics.mcgill.ca \\
email for LH: lavinia.heisenberg@phys.ethz.ch \\
email for JR: jakob.robnik@gmail.com \\
\\
submission date: March 27, 2021\\
\\
Corresponding author: Robert Brandenberger

\date{\today}                                           


\newpage
\normalsize
Extrapolating the solution of Einstein's vacuum field equations to the inside of a black hole horizon, we encounter a curvature singularity \cite{Penrose}. Similarly, extrapolating into the past the solutions of Einstein's equations for a cosmological metric, then, in the presence of a matter source obeying the usual energy conditions, we hit a curvature singularity within a finite time \cite{Hawking}. The black hole and big bang singularities indicate the breakdown of the classical physics effective field theory description of space, time and matter. 

It has long been hoped that quantum approaches to gravity can resolve these singularities. Here, we propose a way to simultaneously resolve black hole and cosmological singularities by the addition of a single object to the effective field theory description of space, time and matter. This object is a S-brane \cite{Sbrane}, a relativistic object which occupies a co-dimension one space-like hypersurface of space-time and carries positive tension but vanishing energy density. This object violates the usual energy conditions and hence enables a resolution of space-time singularities.

Let us first consider the Penrose diagram of a Schwarzschild black hole (see Figure 1). The horizon separates the outside of the black hole (bottom right) from the inside (the top left). Any light ray travelling into the black hole will end up at the singularity (the wavy horizontal curve). According to Einstein's equations, the Weyl curvature diverges as the singular surface is approached. Note that the singularity is a space-like hypersurface.

\begin{figure}
    \includegraphics[scale = 0.5]{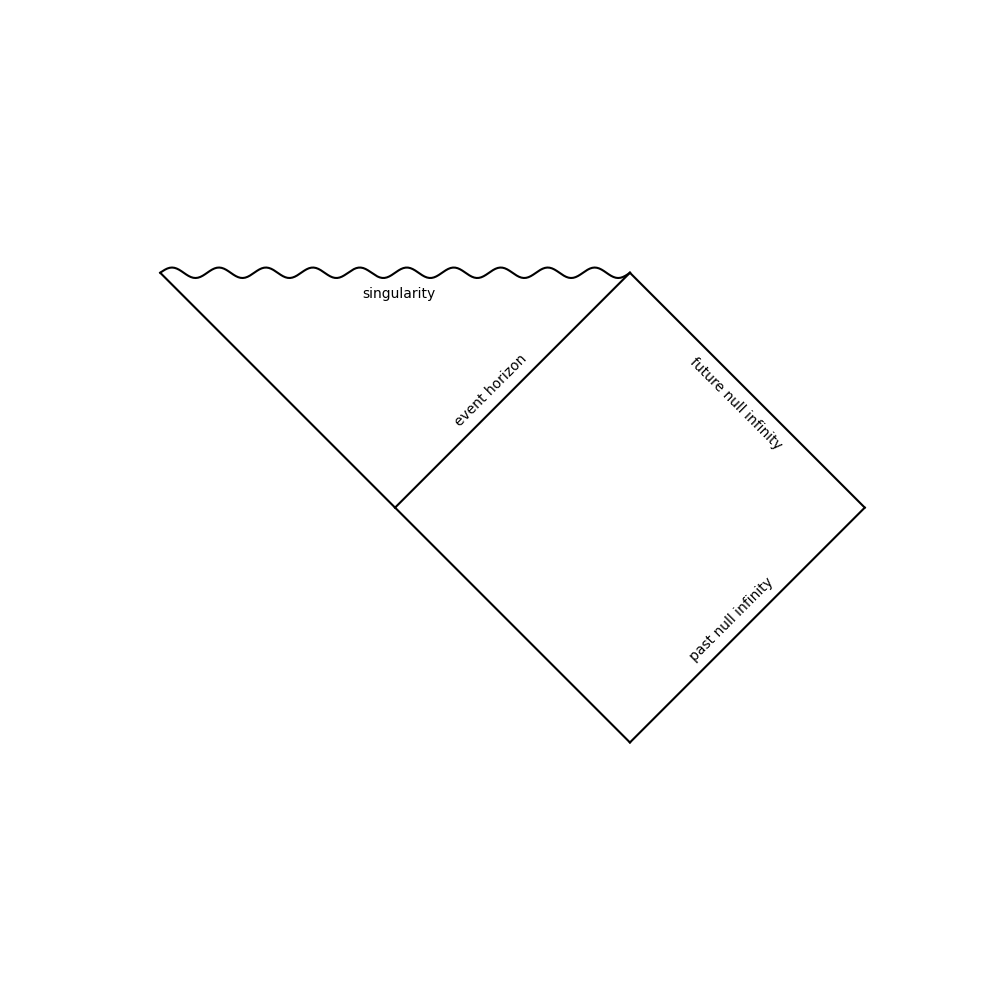}
    \caption{Penrose diagram of the Schwarzschild black hole.}
    \label{fig:conformal}
\end{figure}

A Penrose-like diagram for an expanding cosmology is depicted in Figure 2. As in Figure 1, time increases in the vertical direction and the horizontal direction corresponds to space. The wavy curve at the bottom of the figure represents the Big Bang singularity, a space-like hypersurface.

\begin{figure}
\includegraphics[scale = 0.4]{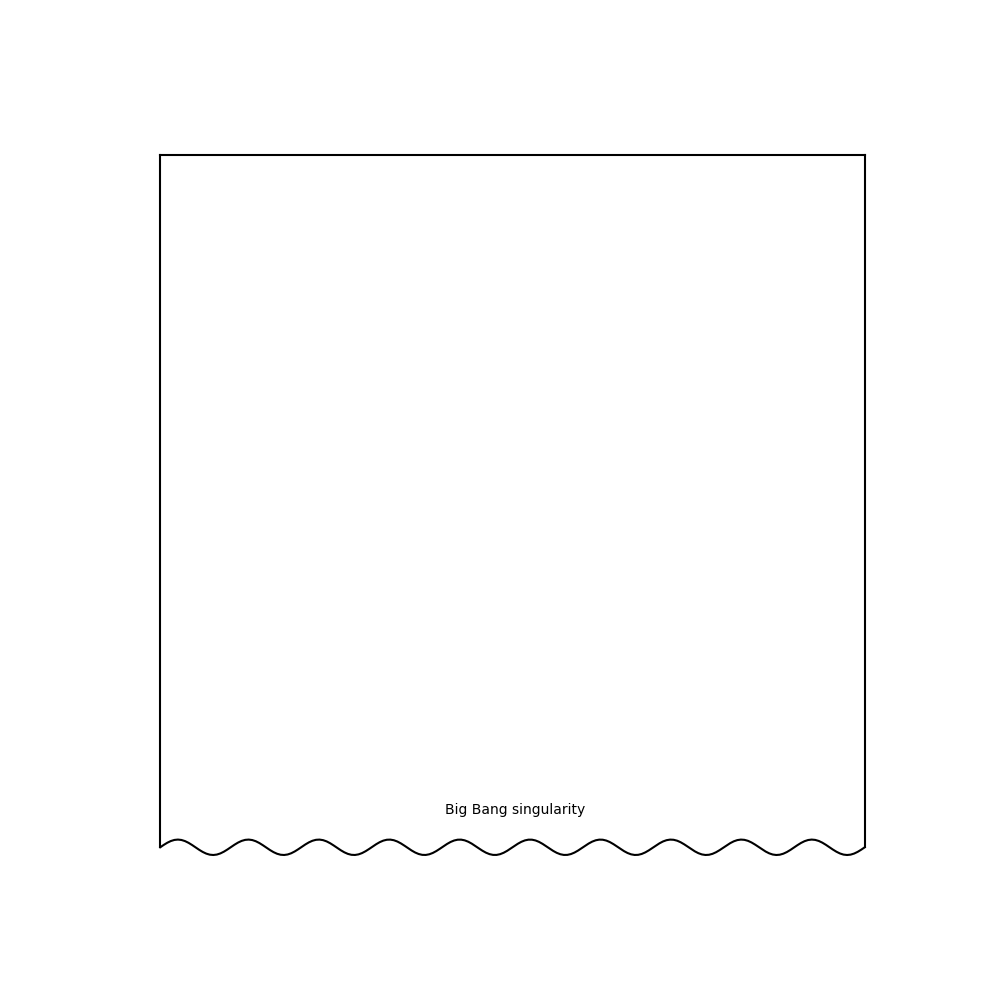}
    \caption{Penrose diagram of an expanding cosmology. The bottom wavy
    line corresponds to the Big Bang singularity.}
    \label{fig:conformal2}
\end{figure}

The coordinates used in Figure 2 were chosen such that it is manifest that it might be possible to glue the Big Bang singularity of Figure 2 to the black hole singularity of Figure 1. In fact, this is the idea of our construction. We cut the black hole space-time on a space-like hypersurface close to the singular surface, a surface where the Weyl curvature takes on a limiting value which is given by new physics beyond General Relativity (we will be inspired by ideas from string theory). Similarly, we cut the cosmological space-time on a space-like hypersurface where the Ricci scalar takes on a limiting value. We then patch the two cut space-times together, as indicated in Figure 3. If we could achieve such a patching, we would be simultaneously resolving both the black hole and Big Bang singularities, obtaining a setup in which an observer could travel into a black hole and emerge in a new universe.

\begin{figure}
    \includegraphics[scale = 0.5]{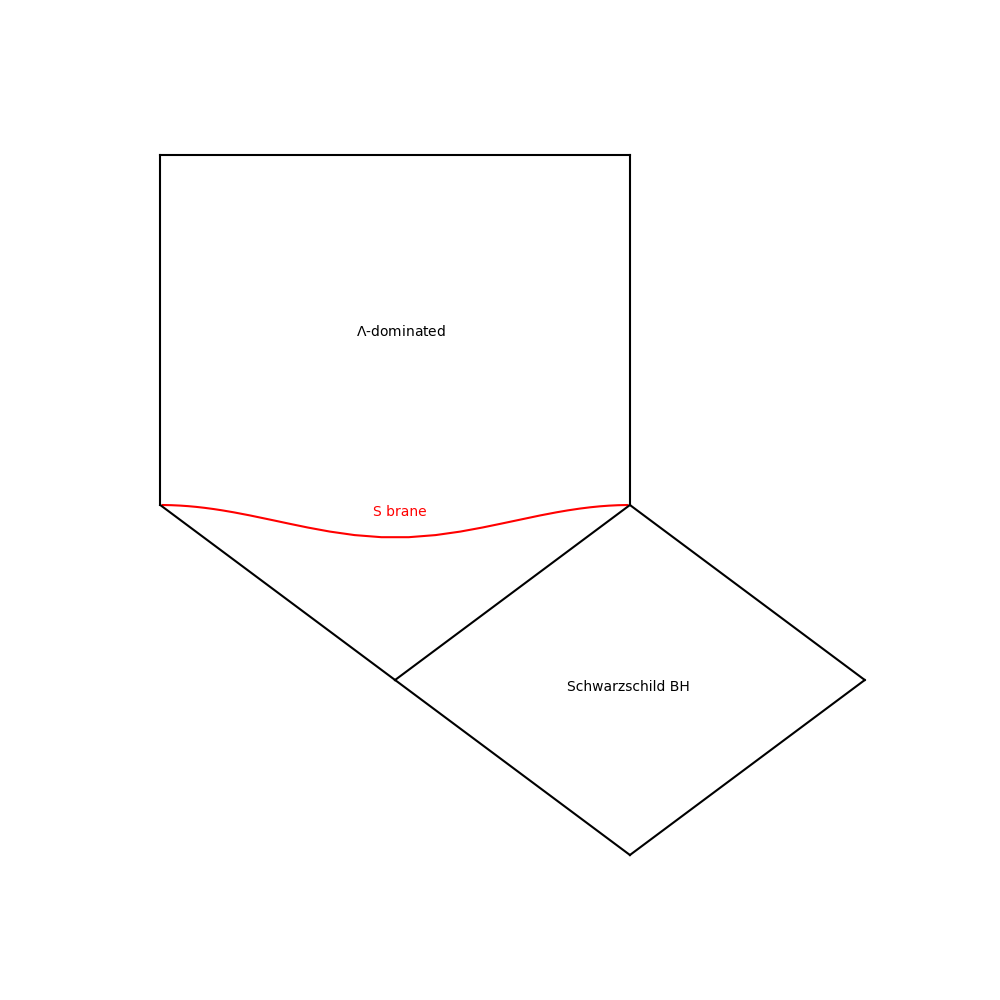}
    \caption{Penrose diagram of the Schwarzschild black hole matched via an S-brane to an expanding de Sitter cosmology.}
    \label{fig:conformal3}
\end{figure}

The idea of obtaining a new universe inside a black hole is an old one (see e.g. \cite{old} for a selection of original ideas). The key challenge is how to obtain a successful gluing between the black hole interior and the expanding cosmology. Making use of the equations of General Relativity coupled to matter obeying the usual energy conditions is not possible, in the same way as it is not possible to transition continuously from a contracting universe to an expanding one. New physics is required in order to obtain a successful transition. The fact that General Relativity leads to singular solutions indicates that if we want to understand space-time in regions of very high curvature, we need to go beyond General Relativity. Superstring theory is a promising approach to go beyond General Relativity. Here, we show that a succesful gluing between a black hole interior and a new universe can naturally be obtained in the context of string theory.

Superstring theory promises, in fact, to unify all forces of nature at the quantum level. The basic objects in string theory are not point particles, but extended objects, namely strings. living in ten space-time dimensions. At low energies, superstring theory reduces to an effective field theory which contains gravity. If we wish to describe the resulting theory at an energy scale $E$, then the effective field theory must include all matter degrees of freedom which are light on the scale $E$. Applied to the inside of a Schwarzschild black hole, this implies that it is consistent to consider pure gravity while the typical energy scale is smaller than the string scale. But as we approach the black hole singularity, then when the typical curvature invariant (e.g. a curvature invariant constructed out of the Weyl tensor) reaches the string scale, the effective field theory must contain an extra matter term. This matter term is confined to a space-like hypersurface of constant curvature \cite{Sbrane}. It is the spacelike cousin of a D-brane, and its energy-momentum tensor is given by a positive tension (i.e. negative pressure) and vanishing energy density (vanishing in the same way that the pressure component perpendicular to a D-brane vanishes). 

Evidently, therefore, the S-brane violates the usual energy conditions and can allow for a transition between the inside of the black hole and the new universe, in the same way as an S-brane was shown to be able to mediate a continous transition between a contracting and an expanding cosmology \cite{Kounnas, Ziwei}.

The starting point of our analysis (see \cite{us} for details) is the following low energy effective action which includes the presence of the S-brane:
\be
    S \, = \, \int dx^4 \bigl[ \sqrt{- \det g} \ \bigg(\frac{1}{2}\mathcal{R} +
     \mathcal{L_M} \bigg) 
     - \delta(r - r_0) \sqrt{- \det q} \ S \bigr].
\ee
where $g$ is the determinant of the space-time metrric, $R$ is the usual Ricci scalar, $q$ is the determinant of the induced metric on the S-brane hypersurface, $\mathcal{L_M}$ is the matter Lagrangian, $S$ is the tension of the S-brane, and $r = r_0$ is the location of the S-brane (recall that inside the horizon the coordinate $r$ is time-like, and $r = 0$ is the location of the Schwarzschild singularity).

We assume that there is no matter present to the past of the S-brane, and we consider the metric to be the usual Schwarzschild metric
\be
ds^2 \, = \, f(r) dt^2 - \frac{1}{f(r)} dr^2 - r^2 d\Omega^2 \, ,
\ee
with $f(r) = 1 - r_s/r$, $r_s$ being the Schwarzschild radius, and $d\Omega^2$ being the metric of the two sphere. To the future of the S-brane we use coordinates corresponding to a homogeneous but anisotropic cosmology
\be
ds^2  = dt^2 - a(t)^2 dx^2 - b(t)^2 d\Omega^2 \, ,
\ee
where $a(t)$ and $b(t)$ are the two scale factors. In these coordinates, the location of the S-brane is some time $t = t_0$.

The initial values of the scale factors and their first derivatives can then be determined by the Israel matching conditions \cite{Israel}: the induced metric on the S-brane computed from both the past and the future must be the same, and the extrinsic curvature jumps by an amount given by the S-brane tension. It can then easily be seen that for a sufficiently large S-brane tension the universe to the future of the S-brane is an expanding cosmology. The matching conditions yield
\begin{align}
    \frac{{\dot{a}}}{a} + \frac{\sqrt{g} f'}{2 f} & = S ,\\
    \frac{{\dot{b}}}{b} + \frac{\sqrt{g}}{r_0} & = S \, ,
\end{align}
where a prime denotes the derivative with respect to $r$, and all quantities are evaluated on the matching surface. As is evident, a positive S-brane tension is required in order to get expansion to the future of the S-brane.

The cosmological evolution to the future of the S-brane depends on the matter content. Matter is produced by the decay of the S-brane, as studied in \cite{Ziwei}. Note that the spatial sections are not isotropic. They have the topology of $R \times S^2$ which they inherit from the topology of the black hole space-time to the interior of the horizon. In order to obtain an isotropic late time cosmology we need to assume that the decay of the S-brane produces matter which acts for some time as an effective cosmological constant. While it is very difficult to obtain such matter from string theory at the level of an effective perturbative field theory treatment, our analysis uses intrinsically non-perturbative concepts, and it is possible that a finite duration quasi de-Sitter phase can be obtained as discussed in \cite{coherent}.

In our setup the famous black hole information loss problem is naturally resolved since the information which falls into the black hole can travel into the new universe. Effectively, the state of the Hawking radiation measured at future null infinity (outside of the black hole) is entangled with the state in the new universe.

To summarize, we have provided a simple and well motivated construction by which both black hole and cosmological singularities are resolved. According to this scenario, the inside of a black hole harbors a new universe. An adventurous traveller venturing into a black hole will emerge in a new universe. In this way, information which falls into a black hole does not get lost but emerges in the new universe. It is also possible to study how small amplitude space-time fluctuations emanating from past infinity outside of the black hole propagate through the horizon and through the S-brane into the new universe \cite{us}.

 
{\tiny {\bf Acknowledgments}. LH is supported by funding from the European Research Council (ERC) under the European Unions Horizon 2020 research and innovation programme grant agreement No 801781 and by the Swiss National Science Foundation grant 179740. JR acknowledges Ad futura Slovenia for support of his MSc studies at the ETH Zurich. The research at McGill is supported in part by funds from NSERC and from the Canada Research Chair program. RB is grateful to the Institute for Theoretical Physics and the Institute for Particle Physics and Astrophysics of the ETH Zurich for hospitality.}

\end{document}